 \documentclass{article}
   \usepackage{graphicx}
\usepackage{amssymb}
\usepackage{amsthm}
\usepackage{amsmath}
\usepackage{multirow}
\newcommand{\secref}[1]{Section~\ref{#1}}
\newcommand{\tabref}[1]{Table~\ref{#1}}
\newcommand{\figref}[1]{Figure~\ref{#1}}

\begin{document}
\title{Modeling Online Reviews 
with Multi-grain Topic Models}

\author{
 Ivan Titov\footnote{This work was done while at Google Inc.}\\
{University of Geneva}\\
{7 route de Drize}\\
{1227 Carouge, Switzerland}\\
\texttt{ivan.titov@cui.unige.ch}
\and
Ryan McDonald\\
{Google Inc.}\\
{76 Ninth Avenue}\\
{New York, NY 10011}\\
\texttt{ryanmcd@google.com}
}

\maketitle
\begin{abstract}
In this paper we present a novel framework for extracting the
ratable aspects of objects from online user reviews.
Extracting such aspects is an important challenge in
automatically mining product opinions from the web
and in generating opinion-based summaries of user reviews \cite{liu2004a,liu2004b,carenini2005,gamon2005,popescu2005,zhuang2006,mei2007}.
Our models are based on extensions to standard topic
modeling methods such as LDA and PLSA to induce multi-grain topics.
We argue that multi-grain models are more appropriate for
our task since standard models
tend to produce topics that correspond to global properties
of objects (e.g., the brand of a product type) rather
than the aspects of an object that tend to be rated
by a user. The models we present not only extract ratable aspects,
but also cluster them into coherent topics, e.g., \textit{waitress} and \textit{bartender}
are part of the same topic \textit{staff} for restaurants.
This differentiates it from much of the previous work which extracts
aspects through term frequency analysis with minimal clustering. We evaluate the
multi-grain models both qualitatively and quantitatively to show that
they improve significantly upon standard topic models.
\end{abstract}

\section{Introduction}

The amount of Web 2.0 content is expanding rapidly.
Due to its source, this content is inherently noisy.
However, UI tools often allow for at least some
minimal labeling, such as topics in blogs, numerical product ratings in
user reviews and helpfulness rankings in online discussion forums.
This unique mix has led to the development of tailored
mining and retrieval algorithms for such content \cite{liu2004a,fujimura2005,ounis2006}.

In this study we focus on online user reviews that have
been provided for products or services, e.g., electronics, hotels and restaurants.
The most studied problem in this domain is sentiment and opinion classification.
This is the task of classifying a text
as being either subjective or objective, or with having positive,
negative or neutral sentiment \cite{wiebe2000,pang2002,turney2002}. However, the sentiment
of online user reviews is often provided by the user. As such,
a more interesting problem is to adapt
sentiment classifiers to blogs and discussion forums to extract
additional opinions of products and services \cite{ounis2006,mei2007}.

Recently, there has been a focus on systems that produce fine-grained sentiment analysis
of user reviews \cite{liu2004b,popescu2005,carenini2006,zhuang2006}.
As an example,
consider hotel reviews. A standard hotel review will probably discuss such aspects
of the hotel like cleanliness, rooms, location, staff, dining experience,
business services, amenities etc. Similarly, a review for a Mp3 player is likely
to discuss aspects like sound quality, battery life, user interface,
appearance etc. Readers are often  interested
not only in the general sentiment towards an object, but also in a detailed opinion analysis for
each these aspects. For instance, a couple on their honeymoon 
are probably not interested in quality of the Internet connection at a hotel,
whereas this aspect can be of a primary importance for a manager on a business trip.

These considerations underline
a need for models that automatically detect aspects discussed in an arbitrary fragment of
a review and predict the sentiment of the reviewer towards these aspects. If
such a  model were available it would be possible to systematically generate a list of 
sentiment ratings for each aspect, and, at the same time, to extract textual evidence from the reviews
supporting each of these ratings. Such a model would have many uses. The example above where users search for
products or services based on a set of critical criteria is one such application.
A second application would be a mining tool for companies that
want fine-grained results for tracking online opinions of their products.
Another application could be Zagat\footnote{http://www.zagat.com} or
TripAdvisor\footnote{http://www.tripadvisor.com} style aspect-based opinion summarizations 
for a wide range of services beyond just restaurants and hotels.

Fine-grained sentiment systems typically solve the task in two phases.
The first phase attempts to extract the aspects of an object that users frequently
rate \cite{liu2004a,carenini2005}. The second phase uses standard
techniques to classify and aggregate sentiment over each of these aspects \cite{liu2004b,carenini2006}.
In this paper we focus on improved models for the first phase -- ratable aspect extraction from
user reviews. In particular, we focus on unsupervised models for extracting
these aspects. The
model we describe can extend both Probabilistic Latent Semantic Analysis \cite{hofmann2001}
and Latent Dirichlet Allocation (LDA) \cite{blei2003} --
both of which are state-of-the-art topic models. We start by showing that standard
topic modeling methods, such as LDA and PLSA,
do not model the appropriate aspects of user reviews. In particular,
these models tend to build topics that globally classify terms into
product instances (e.g., Creative Labs Mp3 players versus iPods,
or New York versus Paris Hotels). To combat this we extend
both PLSA and LDA to induce multi-grain topic models. Specifically,
we allow the models to generate terms from either a global topic
or a local topic that is chosen based on a sliding window context over the
text. The local topics more faithfully model aspects that
are rated throughout the review corpus. Furthermore, the number of
quality topics is drastically improved over standard topic models that
have a tendency to produce many useless topics in addition to a
number of coherent topics.


We evaluate the models both qualitatively and quantitatively.
For the qualitative analysis we present a number of topics
generated by both standard topic models and our new multi-grained
topic models to show that the multi-grain topics are both more coherent
as well as better correlated with ratable aspects of an object. For the quantitative analysis
we will show that the topics generated from the multi-grained topic
model can significantly improve multi-aspect ranking \cite{snyder2007}, which
attempts to rate the sentiment of individual aspects from the text of
user reviews in a supervised setting.

The rest of the paper is structured as follows. 
\secref{SectTopicModels} begins with a review of the standard topic modeling approaches,
PLSA and LDA, and a discussion of their applicability to extracting ratable aspects
of products and services. In the rest of the section we 
introduce a multi-grain model as a way to address the discovered limitations of
PLSA and LDA. \secref{SectInference} describe an inference algorithm for the  multi-grain model. 
In \secref{SectExperims} we provide an
empirical evaluation of the proposed method. We conclude in \secref{SecRelatedWork} with an
examination of related work.

Throughout this paper we use the term \textit{aspect} to denote properties of an object that are
rated by a reviewer. Other terms in the literature include \textit{features}
and \textit{dimensions}, but we opted for \textit{aspects} due to ambiguity
in the use of alternatives.

\section{Unsupervised Topic Modeling}
\label{SectTopicModels} 
As discussed in the preceding section, our goal is to provide a method for extracting ratable
aspects from reviews without any human supervision. Therefore, it is natural to
use generative models of documents, which represent document as mixtures of
latent topics, as a basis for our approach. In this section
we will consider applicability of the most standard methods for unsupervised
modeling of documents, Probabilistic
Latent Semantic Analysis, PLSA~\cite{hofmann2001}  and 
Latent Dirichlet Allocation, LDA~\cite{blei2003} to the considered problem. This analysis will allow us to  
recognize limitations of these models in the context of the considered problem and to propose a new model, Multi-grain LDA,
which is aimed to overcome these limitations. 

\subsection{PLSA \& LDA}
\label{SectLda}

Unsupervised topic modeling has been an area of  active research since the PLSA
method was proposed in~\cite{hofmann2001} as a probabilistic variant of the LSA
method~\cite{deerwester90}, the approach widely used in information  retrieval to
perform dimensionality reduction of documents.
PLSA uses the aspect model~\cite{saul97} to define a generative model of a document. 
It assumes that the document is generated using a mixture of
$K$ topics, where the mixture coefficients are chosen individually for each document.
The model is defined by parameters $\varphi$, $\theta$ and $\rho$, where
$\varphi_z$ is the distribution $P(w|z)$ of words in  latent topic $z$, $\theta_d$ is 
the distribution $P(z|d)$ of topics in  document $d$ and $\rho_d$ is the probability of choosing
document $d$, i.e. $P(d)$. Then, generation of a word in this model is defined as follows:
\begin{itemize}
\item choose document $d \sim \rho$,
\item choose topic $z \sim \theta_d$,
\item choose word $z \sim \varphi_z$.
\end{itemize}
The probability of the observed word-document pair $(d,w)$ can be obtained
by marginalization over latent topics
\begin{eqnarray}
\label{ExprPLSA}
\nonumber
P(d,w) = \rho(d) \sum_z {\theta_d(z) \varphi_{z}(w)}.
\end{eqnarray}

The  Expectation Maximization (EM) algorithm~\cite{dempster77} is used to
calculate maximum likelihood estimates of the parameters.
This will lead to
$\rho(d)$ being proportional to the length of document $d$. As a result, the interesting parts
of the model are the distributions of words in latent topics $\varphi$, and $\theta$,
the distributions of topics in each document. The number of parameters grows linear with
the size of the corpus which leads to  overfitting. A regularized
version of the EM algorithm, Tempered EM (TEM)~\cite{pereira93}, is normally used in practice.  

Along with the need to combat overfitting by using appropriately chosen
regularization parameters, the main drawback of the PLSA
method is that it is inherently transductive, i.e., there is no direct way to
apply the learned model to new documents. In PLSA each document $d$ in the collection 
is represented as a mixture of topics with mixture coefficients $\theta_{d}$, but
it does not define such representation for documents outside the
collection. The hierarchical Bayesian LDA model proposed in~\cite{blei2003} solves both of these problems
by defining a generative model for distributions $\theta_{d}$. 

In LDA, generation of a collection is  started by sampling a word
distribution $\varphi_z$ from a prior Dirichlet distribution $Dir(\beta)$ for each latent topic.
Then each document $d$ is generated as follows:
\begin{itemize}
\item choose  distribution of topics  $\theta_d \sim Dir(\alpha)$
\item for each word $i$ in document $d$
    \begin{itemize}
        \item choose topic $z_{d,i} \sim \theta_d$,
        \item choose word $w_{d,i} \sim \varphi_{z_{d,i}}$.
    \end{itemize}
\end{itemize}

\begin{figure}
\begin{center}
	\begin{minipage}{1.6in}
		\ \\\ \\\ \\\ \\\ \\
	\includegraphics[width=1.4in]{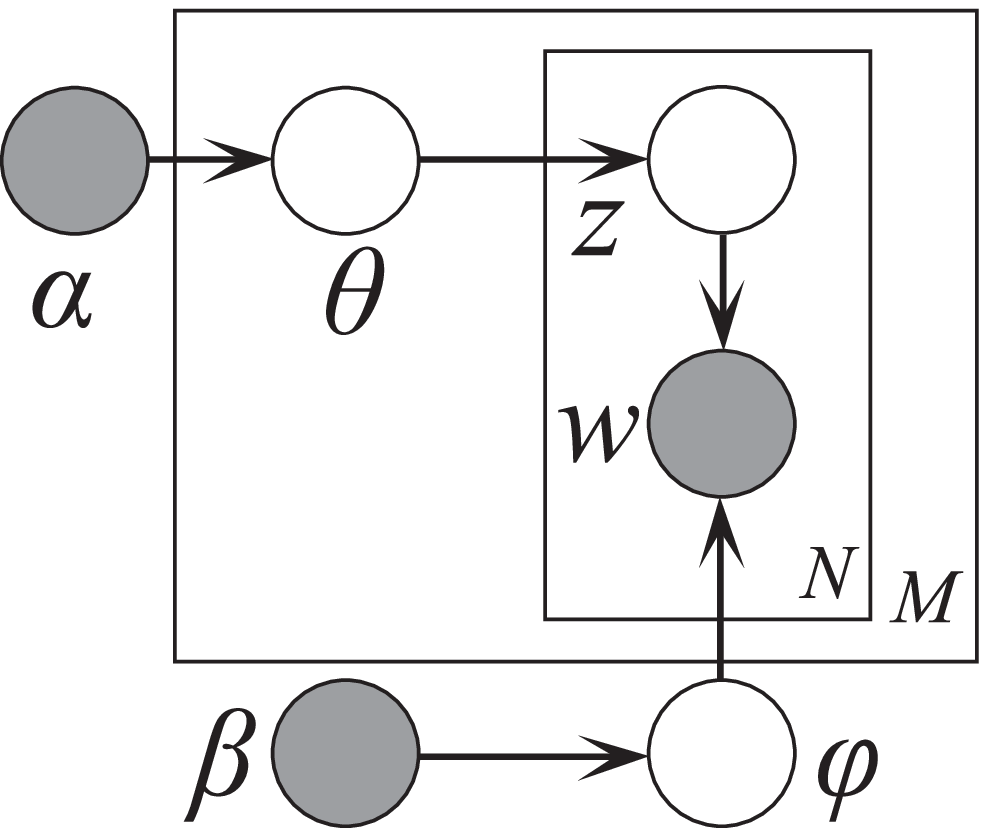}
	\end{minipage}
	\begin{minipage}{1.6in}
	\includegraphics[width=1.6in]{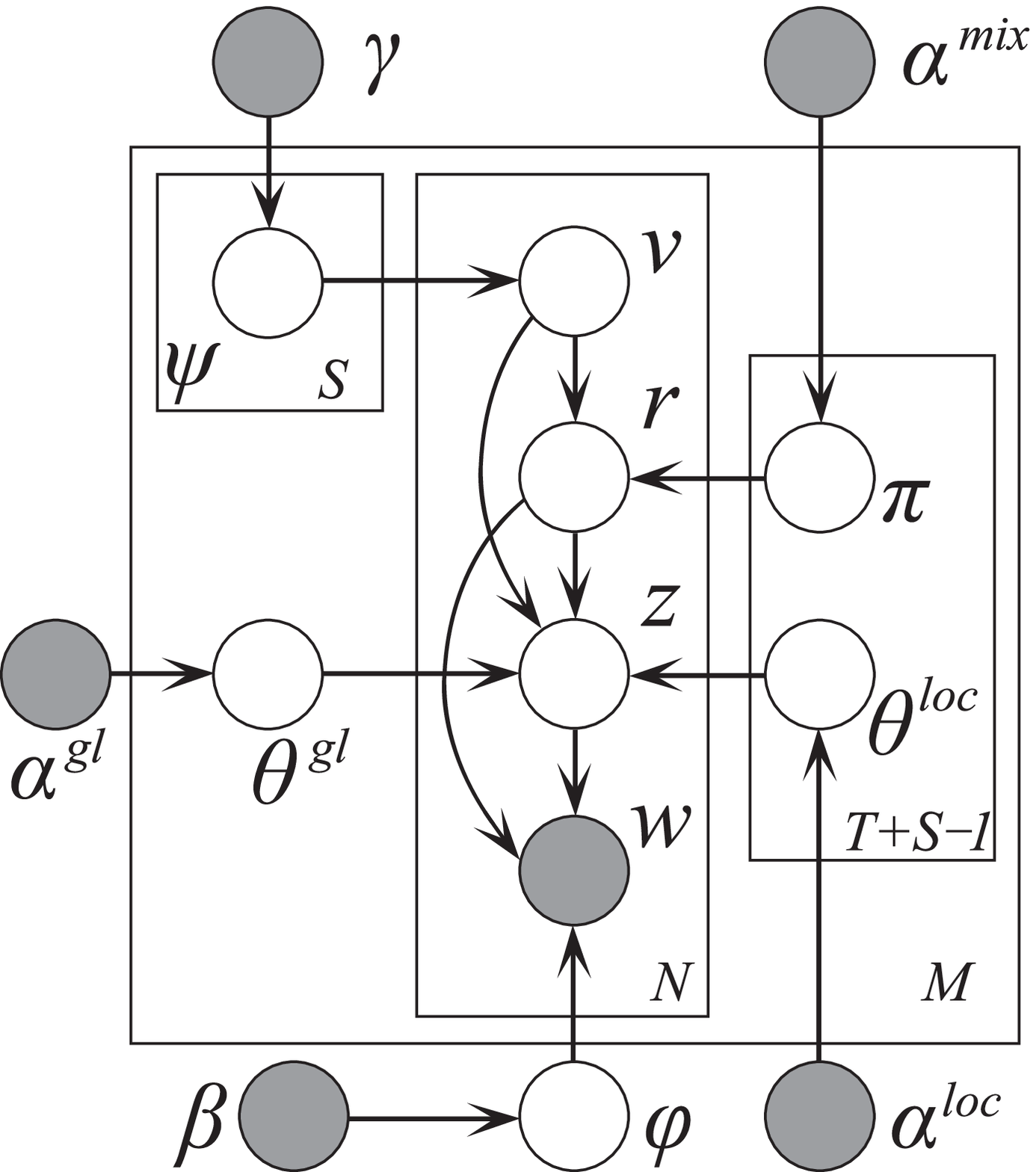}
	\end{minipage}
	\ \\\ \\
	\begin{minipage}{1.6in}
		\begin{center}(a)\end{center}
	\end{minipage}
	\begin{minipage}{1.6in}
		\begin{center}(b)\end{center}
	\end{minipage}
\end{center}
\caption{(a) LDA model. (b) MG-LDA model.}
\label{FigLda}
\end{figure}

The model is represented in \figref{FigLda}a using the standard graphical
model notation. LDA has only two parameters, $\alpha$ and $\beta$,\footnote{Usually 
the symmetrical Dirichlet distribution
$Dir(a)=\frac{1}{B(a)}\prod_i{x_{i}^{a - 1}}$ is used for both of these
priors,  which implies that parameters $\alpha$ and $\beta$ are both scalars.}
which prevents it from overfitting. Unfortunately exact inference in such model
is intractable and various approximations have been considered~\cite{blei2003,minka02,griffiths04}. Originally, 
the variational EM approach 
was proposed in~\cite{blei2003}, which
instead of generating $\varphi$ from  Dirichlet priors, 
a point estimates of distributions $\varphi$ are used and
approximate inference in the resulting model is performed using  variational techniques. The
number of parameters in this empirical Bayes model is still not directly dependent on the number of
documents and, therefore, the model is not expected to suffer from overfitting. 
Another approach is to use  a Markov chain Monte Carlo algorithm for
inference with LDA, as proposed in~\cite{griffiths04}. 
In section~\ref{SectInference} we will describe a modification of this sampling method for 
the proposed Multi-grain LDA model. 

Both LDA and PLSA methods use the bag-of-words representation of documents,  
therefore they can only explore co-occurrences at  the document level. 
This is fine as long as the goal is to represent an overall topic of the
document, but our goal is  different: extracting ratable aspects.
The main topic of all the reviews for a particular item is virtually the same: a review
of this item. Therefore, when such topic modeling methods are applied to a
collection of reviews for different items, they infer topics
corresponding to distinguishing properties of these items. E.g. when applied to a collection of
hotel reviews these models are likely to infer topics: \textit{hotels in France}, \textit{New
York hotels}, \textit{youth hostels}, or, similarly, when applied to a collection of Mp3 players' reviews,
these models will infer topics like \textit{reviews of iPod} or \textit{reviews of Creative Zen player}. 
Though these are all valid topics, they do not represent ratable aspects, but
rather define clusterings of the reviewed items into specific types. In further discussion we will
refer to such topics as \textit{global} topics, because they correspond to a global property 
of the object in the review, such as its brand or base of operation.  
Discovering topics that correlate with ratable aspects, such as \textit{cleanliness} and \textit{location}
for hotels, is much more problematic with LDA or PLSA methods.
Most of these topics are present in some way in every review. Therefore, it is difficult to discover them 
by using only co-occurrence  information at the document level.
In this case exceedingly large amounts of training data is needed and as
well as a very large number of topics $K$. Even in this case there is a danger
that the model  will be overflown by very fine-grain global topics or the resulting
topics will be intersection of global topics and ratable aspects, like
\textit{location for hotels in New York}. We will show in \secref{SectExperims} that
this hypothesis is confirmed experimentally.  

One way to address this problem would be to consider co-occurrences at the sentence
level, i.e., apply LDA or PLSA to individual sentences. 
But in this case we will not have a sufficient co-occurrence domain, and it 
is known that LDA and PLSA behave badly when applied to very short
documents. Though this problem can be addressed by explicitly modeling topic
transitions~\cite{blei01,griffiths04b,wang05,wallach06,purver06,gruber2007}, but these
topic n-gram models are considerably more computationally expensive. Also, like
LDA and PLSA,  they will not be able to distinguish between topics corresponding to ratable 
aspects and global topics representing properties of the reviewed item. 
In the following section we will introduce a method which explicitly models both 
types of topics and efficiently infers
ratable aspects from limited amount of training data.

\subsection{MG-LDA}
\label{SectMgLda}
We propose a model called Multi-grain LDA (MG-LDA), which models two distinct types of topics: 
global topics and local topics. As in PLSA and LDA, the distribution of global topics is fixed for a document.
However, the distribution of local topics is allowed to vary across the document. 
A word in the document is sampled either from the mixture of global topics or from
the mixture of local topics specific for the local context of the word.
The hypothesis is that ratable aspects will be captured by local topics and
global topics will capture properties of reviewed items.
For example consider an extract from a review of a London hotel: ``\ldots public transport in London is straightforward, 
the tube station is about an 8 minute walk \ldots
or you can get a bus for \textsterling 1.50''.  It can be viewed as a mixture
of topic \textit{London} shared by the entire review  (words: ``London'', ``tube'', ``\textsterling''),
and the ratable aspect~\textit{location}, specific for the local context of the sentence  
(words: ``transport'', ``walk'', ``bus'').
Local topics are expected to be reused between very different types of items,
whereas global topics will correspond only to particular types of items.
In order to capture only genuine local topics, we allow a large number of global topics,
effectively, creating a bottleneck at the level of local topics. Of course, this bottleneck
is specific to our purposes. Other applications of multi-grain topic models conceivably might
even prefer the bottleneck reversed. Finally, we note that our definition of multi-grain
is simply for two-levels of granularity, \textit{global}
and \textit{local}. However, there is nothing preventing the model described in this section
from extending beyond two levels. One might expect that for other tasks even
more levels of granularity could be beneficial.

We represent
a document as a set of sliding windows, each covering $T$ adjacent sentences within a document. 
Each window $v$ in document $d$ has an associated  distribution over local topics
$\theta^{loc}_{d,v}$ and  a distribution defining preference for local
topics versus global topics $\pi_{d,v}$. A word can be sampled using any window
covering its sentence $s$, where the window is chosen according to a categorical distribution~$\psi_s$. 
Importantly, the fact that the windows overlap, permits to exploit a larger co-occurrence domain.
These simple techniques are capable of modeling local topics without more expensive modeling 
of topics transitions used in~\cite{blei01,griffiths04b,wang05,wallach06,purver06,gruber2007}. 
Introduction of  a symmetrical Dirichlet prior $Dir(\gamma)$ for the distribution $\psi_s$ 
permits to control smoothness of topic transitions in our model.

The formal definition of the model with $K^{gl}$ global and $K^{loc}$ local topics is the
following. First, draw $K^{gl}$ word distributions for global topics $\varphi^{gl}_{z}$
from a Dirichlet prior $Dir(\beta^{gl})$ and $K^{loc}$ word distributions for local topics
$\varphi^{loc}_{z'}$ - from $Dir(\beta^{loc})$. Then, for each document $d$:
\begin{itemize}
\item Choose a distribution of global topics  $\theta^{gl}_d \sim
Dir(\alpha^{gl})$.
\item For each sentence $s$ choose a distribution $\psi_{d,s}(v) \sim Dir(\gamma)$.
\item For each sliding window $v$ 
    \begin{itemize}
        \item choose $\theta^{loc}_{d,v} \sim Dir(\alpha^{loc})$,
        \item choose $\pi_{d,v} \sim Beta(\alpha^{mix})$.
    \end{itemize}

\item For each word $i$ in sentence $s$ of document $d$
    \begin{itemize}
        \item choose window $v_{d,i} \sim \psi_{d,s}$,
        \item choose $r_{d,i} \sim \pi_{d,v_{d,i}}$,
        \item if $r_{d,i} = gl$ choose  global topic $z_{d,i} \sim \theta^{gl}_d$,
        \item if $r_{d,i} = loc$ choose local topic $z_{d,i} \sim  \theta^{loc}_{d,v_{d,i}}$,
        \item choose word $w_{d,i}$ from the word distribution $\varphi^{r_{d,i}}_{z_{d,i}}$.
    \end{itemize}
\end{itemize}
Here, $Beta(\alpha^{mix})$ is a prior Beta distribution for  choosing between local and
global topics. Though  symmetrical Beta distributions can be considered, we  use 
a non-symmetrical one as it permits to regulate preference to either 
global or  local topics by setting 
$\alpha^{mix}_{gl}$ and $\alpha^{mix}_{loc}$ accordingly. 

In \figref{FigLda}b the corresponding graphical model is presented.
As we will show in the following section this model allows for fast approximate
inference with collapsed Gibbs sampling.


\section{Inference with MG-LDA}
\begin{table*}
\centering
\caption{Datasets used for qualitative evaluation.}
\begin{tabular}{|l|c|c|c|c|} \hline
Domain & Reviews & Sentences & Words & Words per review \\\hline  
Mp3 players & 3,872 & 69,986 & 1,596,866 & 412.4\\ 
Hotels  & 32,861  & 264,844 & 4,456,972 & 135.6 \\ 
Restaurants & 32,563 & 136,906 & 2,513,986 & 77.2 \\
\hline\end{tabular}
\label{TabDatasets}
\end{table*}

\label{SectInference}
In this section we will describe a modification  of the inference algorithm
proposed in~\cite{griffiths04}. 
But before starting with our Gibbs sampling algorithm we should note that instead of 
sampling from Dirichlet and Beta priors we could fix 
$\psi_{d,s}$ as a uniform distribution and compute maximum likelihood estimates for $\varphi^r$ and 
$\theta^r$. Such model can be trained by using the EM algorithm 
or the TEM algorithm  and viewed as a generalization of the PLSA aspect model. 


Gibbs sampling is an example of a Markov Chain
Monte Carlo algorithm~\cite{geman84}. It is used to produce a sample from a joint
distribution when only conditional distributions of each variable
can be efficiently computed. In Gibbs sampling, variables are sequentially 
sampled from their distributions conditioned on all other variables in the
model. Such a chain of model states converges to a sample from the joint distribution.
A naive application of this technique to LDA would imply that both assignments
of topics to words $\textbf{z}$  and distributions $\theta$ and $\varphi$ should be
sampled. However, Griffiths and Steyvers~\cite{griffiths04} demonstrated that
an efficient collapsed Gibbs sampler can be constructed, where only assignments
$\textbf{z}$ need to be sampled, whereas the dependency on distributions $\theta$ and $\varphi$
can be integrated out analytically. Though derivation of the collapsed Gibbs sampler for
MG-LDA is similar to the one proposed by Griffiths and Steyvers for LDA, we rederive 
it here for completeness.

In order to perform Gibbs sampling with MG-LDA we need to compute conditional probability
$P(v_{d,i} = v, r_{d,i} = r,$ $ z_{d,i} = z |\textbf{v'}, $ $\textbf{r'}, \textbf{z'}, \textbf{w})$,
where $\textbf{v'}$, $\textbf{r'}$ and $\textbf{z'}$ are vectors of
assignments of sliding windows, context (global or local) and topics for all the
words in the collection except for the considered word at position $i$ in
document $d$. We denote by $\textbf{w}$  a vector of all the words in the collection. We start by
showing how the joint probability of the assignments and the words $P(\textbf{w},
\textbf{v},\textbf{r},\textbf{z}) =$$P(\textbf{w}|
\textbf{r},\textbf{z})P(\textbf{v},\textbf{r},\textbf{z})$  can be evaluated.
By integrating  out $\varphi^{gl}$ and $\varphi^{loc}$ we can obtain the first term:
\begin{eqnarray}
\label{ExprWordProb}
P(\textbf{w}|\textbf{r},\textbf{z}) =\!\!\!
\prod_{r \in \{gl,loc\}} {
\!
\left(
\frac {\Gamma(W\beta^{r})}  {{\Gamma(\beta^{r})}^W}
\right)^{\!\!K^{r}}
\!
\prod_{z = 1}^{K^{r}}
{ 
\frac{ \prod_w{\Gamma(n^{r,z}_{w} + \beta^{r})}}{\Gamma(n^{r,z} + W\beta^r)}
}
},
\end{eqnarray}
where $W$ is the size of the vocabulary,  $n^{gl,z}_{w}$ and $n^{loc,z}_{w}$ are
the numbers of times word $w$ appeared in global and local topic $z$, $n^{gl,z}$
and $n^{loc,z}$ are the total number of words assigned to global or local topic
$z$, and $\Gamma$ is the gamma function. 
To evaluate the second term, we factor it as
$P(\textbf{v},\textbf{r},\textbf{z}) = P(\textbf{v}) P(\textbf{r}|\textbf{v})
P(\textbf{z}|\textbf{r},\textbf{v})$ and compute each of these factors
individually. By integrating out $\psi$ we obtain
\begin{eqnarray}
\label{ExprWinAssignProb}
P(\textbf{v}) =
\left( {
\frac{ \Gamma(T\gamma)}{\Gamma(\gamma)^T}
}\right)^{N_s}
\prod_{d,s}{
\frac
    {\Gamma(n^{d,s}_v + \gamma)}
    {\Gamma(n^{d,s} + T\gamma)}
},
\end{eqnarray}
in which $N_s$ denotes the number of sentences in the collection, 
$n^{d,s}$ denotes the length of sentence $s$ in document $d$, and
$n^{d,s}_v$ is the number of times a word from this sentence is assigned to
window $v$.
Similarly, by integrating out $\pi$ we compute
\begin{eqnarray}
\nonumber
P(\textbf{r}|\textbf{v})=
\left( {
\frac
    {\Gamma(\sum_{r \in \{gl, loc\}}{\alpha^{mix}_r})}
    {\prod_{r \in \{gl, loc\}}{\Gamma(\alpha^{mix}_r)}}
} \right) ^{N_v}
\\
\label{ExprContextAssignProb}
\prod_{d,v} {
\frac
    {\prod_{r \in \{gl,loc\}}{\Gamma(n_{r}^{d,v} + \alpha^{mix}_r)}}
    {\Gamma(n^{d,v} +  \sum_{r \in \{gl, loc\}}{\alpha^{mix}_r})}
},
\end{eqnarray}
in this expression $N_v$ is the total number of  windows in the collection,
$n^{d,v}$ is the number of words assigned to window $v$,
$n_{gl}^{d,v}$ and $n_{loc}^{d,v}$ are the number of times a word from 
window $v$ was assigned to global and to local topics, respectively.  
Finally, we can compute conditional probability of assignments of words to topics
by integrating out~both~$\theta^{gl}$~and~$\theta^{loc}$
\begin{eqnarray}
P(\textbf{z}|\textbf{r},\textbf{v}) =
\left( {
\frac{\Gamma(K^{gl}\alpha^{gl})}{\Gamma(\alpha^{gl})^{K^{gl}}}
} \right)^{D}
\prod_d {
\frac
    {\prod_{z} {\Gamma(n^{d}_{gl,z} + \alpha^{gl})}}
    {\Gamma(n^{d}_{gl} + K^{gl} \alpha^{gl})} 
}
\label{ExprTopicAssignProb}
\left( {
\frac{\Gamma(K^{loc}\alpha^{loc})}{\Gamma(\alpha^{loc})^{K^{loc}}}
} \right)^{N_v}
\prod_{d,v} {
\frac
    {\prod_{z} {\Gamma(n^{d,v}_{loc,z} + \alpha^{loc})}}
    {\Gamma(n^{d,v}_{loc} + K^{loc} \alpha^{loc})} 
},
\end{eqnarray}
here $D$ is the number of documents, $n^{d}_{gl}$ is the number of times a word in
document $d$ was assigned to one of the global topics and $n^{d}_{gl,z}$ is the
number of times a word in this document was assigned to global topic $z$.
Similarly, counts $n^{d,v}_{loc}$ and $n^{d,v}_{loc,z}$ are defined for local topics in
window $v$ in document $d$. Now  the conditional distribution
$P(v_{d,i} = v, r_{d,i} = r, z_{d,i} = z |\textbf{v'}, \textbf{r'}, \textbf{z'},
\textbf{w})$ can be obtained by cancellation of terms in
expressions~(\ref{ExprWordProb}-\ref{ExprTopicAssignProb}).
For global topics we get
\begin{align}
\nonumber
P(v_{d,i} = v,& r_{d,i} = gl, z_{d,i} = z |\textbf{v'}, \textbf{r'},
\textbf{z'},\textbf{w}) \propto 
\frac{n^{gl,z}_{w_{d,i}} + \beta^{gl}} {n^{gl,z} + W \beta^{gl}}
\\
\nonumber
&
\times \frac {n^{d,s}_v + \gamma} {n^{d,s} + T \gamma} \times
\frac {n^{d,v}_{gl} + \alpha^{mix}_{gl}}{n^{d,v} + \sum_{r' \in \{gl,loc\} }{\alpha^{mix}_{r'}}}  \times
\frac {n^{d}_{gl,z} + \alpha^{gl}}{n^d_{gl} + K^{gl} \alpha^{gl}}
,
\end{align}
where $s$ is the sentence in which the word $i$ appears. Here factors correspond
to the probabilities of choosing word $w_{d,i}$, choosing window $v$, choosing
global topics and choosing topic $z$ among global topics. For
local topic the conditional probability is estimated as
\begin{align}
\nonumber
P(v_{d,i} = v, & r_{d,i} = loc, z_{d,i} = z |\textbf{v'}, \textbf{r'},
\textbf{z'},\textbf{w}) \propto 
\frac{n^{loc,z}_{w_{d,i}} + \beta^{loc}} {n^{loc,z} + W \beta^{loc}}
\\
\nonumber
&
\times \frac {n^{d,s}_v + \gamma} {n^{d,s} + T \gamma} \times
\frac {n^{d,v}_{loc} + \alpha^{mix}_{loc}}{n^{d,v} + \sum_{r' \in \{gl,loc\} }{\alpha^{mix}_{r'}}}  \times
\frac {n^{d,v}_{loc,z} + \alpha^{loc}}{n^{d,v}_{loc} + K^{loc} \alpha^{loc}}
.
\end{align}

In both of these expressions counts are computed without taking into account 
assignments of the considered word $w_{d,i}$.  Sampling with such model is 
fast and in practice convergence with MG-LDA and
can be achieved in time similar to that needed for standard LDA implementations.

A sample obtained from such chain can be used to approximate the distribution of
words in topics:
\begin{eqnarray}
\label{ExprTopicModelEstim}
\hat{\varphi}^{r}_{z}(w) \propto n^{r,z}_w + \beta^{r}.
\end{eqnarray}
The distribution of topics in sentence $s$ of document $d$ can be estimated as
follows
\begin{eqnarray}
\label{ExprTopicDistrEstimGl}
\hat{\theta}^{gl}_{d,s}(z)
=
\sum_v{
\frac {n^{d,s}_v + \gamma} {n^{d,s} \!\!+\!\! T \gamma} \! \times \!
\frac {n^{d,v}_{gl} + \alpha^{mix}_{gl}}{n^{d,v} \!\!+\!\! \sum_{r' \in \alpha^{mix}_{r'}}} \!  \times \!
\frac {n^{d}_{gl,z} + \alpha^{gl}}{n^d_{gl}\!\! +\!\! K^{gl} \alpha^{gl}}
},
\end{eqnarray}
\begin{eqnarray}
\label{ExprTopicDistrEstimLoc}
\hat{\theta}^{loc}_{d,s}(z)
=
\!\sum_v{
\frac {n^{d,s}_v + \gamma} {n^{d,s}\!\! +\!\! T \gamma} \! \!\times\!\!
\frac {n^{d,v}_{loc} + \alpha^{mix}_{loc}}{n^{d,v} \!\!+ \!\!\sum_{r' \in \alpha^{mix}_{r'}}} \!\! \times\!\!
\frac {n^{d,v}_{loc,z} + \alpha^{loc}}{n^{d,v}_{loc}\!\! +\!\! K^{loc} \alpha^{loc}}
}.
\end{eqnarray}

As explained in~\cite{griffiths04b}, the problem of the collapsed sampling approach is 
that when computing statistics it is not possible to aggregate over several samples from 
the probabilistic model. It happens because there is no correspondence between 
indices of topics in different samples. Though for large collections one sample is 
generally sufficient, but with small collections such estimates might become very random. 
In all our experiments we used collapsed sampling methods. For smaller
collections maximum likelihood estimation with EM can be used or variational
approximations can be derived~\cite{blei2003}.






\section{Experiments}
\label{SectExperims}

In this section we present qualitative and quantitative experiments.
For the qualitative analysis we show that local
topics inferred by MG-LDA do correspond to ratable aspects. We compare the
quality of topics obtained by MG-LDA with topics discovered by the standard LDA approach. 
For the quantitative analysis we show that the topics generated from the
multi-grain models can significantly improve multi-aspect ranking.

\subsection{Qualitative Experiments}

\subsubsection{Data}

To perform qualitative experiments we used a subset of reviews for Mp3 players from Google Product 
Search\footnote{http://www.google.com/products}
and subsets  of reviews of hotels and restaurants from Google Local 
Search.\footnote{http://local.google.com} These reviews are either entered by
users directly through Google, or are taken from review feeds provided
by CNet.com, Yelp.com, CitySearch.com, amongst others.
All the datasets were tokenized and sentence split.
Properties of these 3 datasets are presented in
table~\ref{TabDatasets}.
Before applying the topic models we removed punctuation and also removed stop words using the standard
list of stop words.\footnote{http://www.dcs.gla.ac.uk/idom/ir\_resources/linguistic\_utils/ stop\_words}

\subsubsection{Experiments and Results}

\begin{table*}
\centering
\caption{Top words from MG-LDA and LDA topics for Mp3 players' reviews.}
\begin{scriptsize}
\begin{tabular}{|l|l|l|} \hline
   & label & top words \\ \hline
& sound quality & 
sound quality headphones volume bass earphones  good settings ear rock excellent\\
& features &
games features clock contacts calendar alarm notes game quiz feature extras solitaire\\
& connection with PC&
usb pc windows port transfer computer mac software cable xp connection plug firewire\\
& tech. problems & 
reset noise backlight slow freeze turn remove playing icon creates hot cause disconnect \\
MG-LDA& appearance & 
case pocket silver screen plastic clip easily small blue black light white belt cover \\
local & controls &
button play track menu song buttons volume album tracks artist screen press select \\
(all topics)& battery &
battery hours life batteries charge aaa rechargeable time power lasts hour charged \\
& accessories &
usb cable headphones adapter remote plug power charger included case firewire \\
& managing files &
files software music computer transfer windows media cd pc drag drop file using \\
& radio/recording & 
radio fm voice recording record recorder audio mp3 microphone  wma formats \\ 
\hline
& iPod &
ipod music apple songs use mini very just itunes like easy great time new buy really \\
MG-LDA & Creative Zen & 
zen creative micro touch xtra pad nomad waiting deleted labs nx sensitive 5gb eax  \\
global & Sony Walkman &
sony walkman memory stick sonicstage players atrac3 mb atrac far software format \\
& video players &
video screen videos device photos tv archos pictures camera movies dvd files view  \\
&  support  &
player product did just bought unit got buy work \$ problem support  time months \\
\hline
 & iPod &
ipod music songs itunes mini apple battery use very computer easy time just song \\
& Creative & 
creative nomad zen xtra jukebox eax labs concert effects nx 60gb experience lyrics \\
LDA & memory/battery & 
card memory cards sd flash batteries lyra battery aa slot compact extra mmc 32mb\\
(out of 40) & radio/recording& 
radio fm recording record device audio voice unit battery features usb recorder\\
& controls&
button menu track play volume buttons player song tracks press mode screen settings \\
& opinion   &
points reviews review negative bad general none comments good please content aware \\
&   -        & 
player very use mp3 good sound battery great easy songs quality like just music \\
\hline\end{tabular}
\end{scriptsize}
\label{TabMp3}
\end{table*}

We used the Gibbs sampling algorithm both for MG-LDA and LDA, and ran the chain for
800 iterations  to produce a sample for each of the experiments.  Distributions of words in
each topic were then estimated as in~(\ref{ExprTopicModelEstim}).
The sliding windows were chosen to cover 3 sentences for all the experiments.
Coarse tuning of parameters of the prior distributions was performed
both for the MG-LDA and LDA models.  
We varied the number of topics in LDA and  the number 
of local and global  topics in MG-LDA. 
Quality of local topics  for MG-LDA did not seem to be influenced by the number of global topics $K^{gl}$
as long as $K^{gl}$ exceeded the number of local topics $K^{loc}$ by factor of 2. 
For Mp3 and hotel reviews' datasets, when increasing  $K^{loc}$  most of the local topics  
represented ratable aspects until a point when further
increase of $K^{loc}$ was started to produce mostly non-meaningful topics. For LDA we
selected the topic number corresponding to the largest number of discovered
ratable aspects. In this way our comparison was as fair to LDA as possible.

Top words for the discovered local topics and for some of the global topics of MG-LDA models are presented in
\tabref{TabMp3}\ -\ \tabref{TabRests}, one topic  per line, along with selected topics from the LDA 
models.\footnote{Though
we did not remove numbers from the datasets before applying the topic models, 
we removed them from the tables of results to improve readability.}
We manually assigned  labels to coherent topics to reflect our 
interpretation of their meaning.
Note that the MG-LDA local topics in 
\tabref{TabMp3} and \tabref{TabHotels} represent the entire set of local topics used in
MG-LDA models. 
In the meantime, for the LDA topics we selected only the coherent topics which
captured ratable aspects and additionally a number of example topics to show typical LDA
topics. Global topics of MG-LDA are not supposed to capture ratable
aspects and they are not of primary interest in these experiments. In the tables we presented only 
typical MG-LDA global topics and any global
topics which, contrary to our expectations, discovered ratable aspects.

For the reviews of Mp3 players we present results of the MG-LDA model with 10
local and 30 global topics. All 10 local topics seem to correspond to ratable
aspects.  Furthermore, the majority of global topics represent brands of Mp3 players or 
additional categorizations of players such as those with video capability.
The only genuine ratable aspect in the set of
global topics is \textit{support}. Though not entirely clear, but presence of \textit{support}
topic in the list of global  topics might be explained by the considerable number of
reviews in the dataset focused almost entirely on problems with technical
support. The LDA model had 40 topics and only 4 of them (\textit{memory/battery}, \textit{radio/recording}, \textit{controls} and possibly \textit{opinion})
corresponded to ratable aspects. And even these 4 topics are
of relatively low quality.
Though mixing related topics \textit{radio} and \textit{recording} is probably appropriate, combining
concepts \textit{memory} and \textit{battery} is clearly undesirable. Also top
words for LDA topics contain entries corresponding to player properties or brands
(as \textit{lyra} in \textit{memory/battery}), or not so related words (as \textit{battery} and \textit{unit} in \textit{radio/recording}).
In words beyond top 10 this happens for LDA much more frequently than for
MG-LDA.
Typical topics of the LDA model either seem not semantically coherent (as the
last topic in \tabref{TabMp3}) or represent player brands or types. 

\begin{table*}[t]
\centering
\caption{Top words from MG-LDA and LDA topics for hotel reviews.}
\begin{scriptsize}
\begin{tabular}{|l|l|l|} \hline
   & label & top words \\ \hline
&amenities &
coffee microwave fridge tv ice room refrigerator machine kitchen maker iron dryer\\
&food and drink&
food restaurant bar good dinner service breakfast ate eat drinks menu buffet meal\\
& noise/conditioning & 
air noise door room hear open night conditioning loud window noisy doors windows\\
& bathroom &
shower water bathroom hot towels toilet tub bath sink pressure soap shampoo \\
& breakfast &
breakfast coffee continental morning fruit fresh buffet included free hot
juice\\
 & spa &
pool area hot tub indoor nice swimming outdoor fitness spa heated use kids\\
MG-LDA & parking &
parking car park lot valet garage free street parked rental cars spaces space\\
local & staff &
staff friendly helpful very desk extremely help directions courteous concierge\\
(all topics)& Internet &
internet free access wireless use lobby high computer available speed business\\
& getting there &
airport shuttle minutes bus took taxi train hour ride station cab driver line\\
& check in & 
early check morning arrived late hours pm ready day hour flight wait\\
& smells/stains & 
room smoking bathroom smoke carpet wall smell walls light ceiling dirty \\
& comfort &
room bed beds bathroom comfortable large size tv king small double bedroom\\
& location & 
walk walking restaurants distance street away close location shopping shops\\
& pricing &
\$ night rate price paid worth pay cost charge extra day  fee parking \\
\hline
MG-LDA & beach resorts & 
beach ocean view hilton balcony resort ritz island head club pool oceanfront \\
global & Las Vegas &
vegas strip casino las rock hard station palace pool circus renaissance \\
\hline
 & beach resorts &
beach great pool very place ocean stay view just nice stayed clean beautiful \\
& Las Vegas &
vegas strip great casino \$ good hotel food las rock room very pool nice \\
& smells/stains &
room did smoking bed night stay got went like desk smoke non-smoking smell\\
& getting there &
airport hotel shuttle bus very minutes flight hour free did taxi train car\\
& breakfast &
breakfast coffee fruit room juice fresh eggs continental very toast morning\\
LDA & location & 
hotel rooms very centre situated well location excellent city comfortable good\\
(out of 45) & pricing & 
card credit \$ charged hotel night room charge money deposit stay pay cash did \\
& front desk &
room hotel told desk did manager asked said service called stay rooms \\
& noise & 
room very hotel night noise did hear sleep bed door stay floor time just like\\
& opinion &
hotel best stay hotels stayed reviews service great time really just say rooms\\
& cleanliness   & 
hotel room dirty stay bathroom rooms like place carpet old very worst bed \\
&  -  & 
motel rooms nice hotel like place stay parking price \$ santa stayed good \\
\hline\end{tabular}
\end{scriptsize}
\label{TabHotels}
\end{table*}

For the hotels reviews we present results of the MG-LDA model with 15 local topics
and 45 global topics and results of the  LDA model with 45 topics.
Again, top words for all the MG-LDA local topics are given
in \tabref{TabHotels}. Only 9 topics out of 45 LDA topics corresponded to
ratable aspects and these are shown in the table. Also, as with the Mp3
player reviews, we chose 3 typical LDA topics (\textit{beach resorts}, \textit{Las
Vegas} and the last, not coherent, topic).
All the local topics of MG-LDA again reflect ratable
aspects and no global topics seem to capture any ratable aspects. All the
global topics of MG-LDA appear to correspond to hotel types and locations, such
as beach resorts or hotels in Las Vegas, though some global topics are not
semantically coherent. Most of LDA topics are similar to MG-LDA global
topics.
We should note that as with the Mp3 reviews, increasing number of topics for LDA 
beyond 45 did not bring any more
topics corresponding to ratable aspects.

\begin{table*}[t]
\centering
\caption{Top words from MG-LDA and LDA topics for restaurant reviews.}
\begin{scriptsize}
\begin{tabular}{|l|l|l|} \hline
& label & top words \\ \hline
      &service &  
did table said order told minutes got asked waiter waitress came ordered took  \\
MG-LDA      & atmosphere &
bar nice beer music people fun drinks sit outside enjoy watch live cool drink night  \\
local      & location &
restaurant location street right located little downtown parking away cafe near \\
      & previous experience & 
times years time going used eaten ago new eat year visit old couple gone week \\
      & decor &
restaurant dining room decor atmosphere bar tables feel small like old inside cozy  \\
\hline
\multirow{3}{*}{LDA} & ice cream & 
ice cream  soft-serve favorite such butter banana little   chocolate peanut popular \\
& waiting & 
wait long place worth time lunch great line best busy parking always waiting table \\
& service &
food did service table time restaurant minutes never ordered order said just told\\
\hline\end{tabular}
\end{scriptsize}
\label{TabRests}
\end{table*}

The dataset of restaurant reviews appeared to be challenging for both of
the models.  Both MG-LDA and LDA models managed to capture only few ratable
aspects. We present topics corresponding to ratable aspects from 
the MG-LDA model with 20 local and 50 global topics  and from the LDA model with 60
topics. One problem with this dataset is that restaurant reviews are generally short, average
length of a review is 4.2 sentences. Also these results can probably be explained 
by observing the fact that
the majority of natural ratable aspects are specific for a type of restaurants.
E.g., appropriate ratable aspects for Italian restaurants could be \textit{pizza} and
\textit{pasta}, whereas for Japanese restaurants they are probably \textit{sushi} and \textit{noodles}.
We could imagine generic categories like \textit{meat dishes} and \textit{fish dishes} but
they are unlikely to be revealed by any unsupervised model as the overlap
in the vocabulary describing these aspects in  different cuisines is small.\footnote{We ran
preliminary experiments which suggested that MG-LDA is able to infer appropriate
ratable aspects if applied to a set of reviews of restaurants with a specific
cuisine. 
E.g. for MG-LDA with 15 local topics applied to the collection of Italian
restaurant reviews, 9 topics corresponded to ratable
dimensions: \textit{wine}, \textit{pizza}, \textit{pasta},
\textit{general food}, 
\textit{location}, 
\textit{service}, \textit{waiting}, 
\textit{value} and
\textit{atmosphere}.  This result agrees with our explanation.}
One approach to address this problem is to attempt hierarchical topic
modeling~\cite{blei04,mccallum07}.

\subsection{Quantitative Experiments}

\subsubsection{Data and Problem Set-up}

Topic models are typically evaluated quantitatively using measures like likelihood on
held-out data \cite{hofmann2001,blei2003,gruber2007}. However, likelihood
does not reflect our actual purpose since we are not trying to predict whether
a new piece of text is likely to be a review of some particular category. Instead
we wish to evaluate how well our learned topics correspond to aspects of an object
that users typically rate.

To accomplish this we will look at the problem of multi-aspect opinion rating \cite{snyder2007}.
In this task a system needs to predict a discrete numeric rating for multiple aspects
of an object. For example, given a restaurant review, a system would predict
on a scale of 1-5 how a user liked the food, service, and decor of the restaurant.
This is a challenging problem since users will use a wide variety of language to describe
each aspect. A user might say ``The X was great'', where X could be ``duck'',
``steak'', ``soup'', each indicating that the food aspect should receive a high rating.
If our topic model identifies a food topic (or topics), then this information
could be used as valuable features when predicting the sentiment of an aspect since it
will inform the classifier which sentences are genuinely about which aspects.

To test this we downloaded 27,564 hotel reviews from TripAdvisor.com.\footnote{http://www.tripadvisor.com}
These reviews are labeled with a rating of 1-5 for a variety of ratable aspects
for hotels. We selected our review set to span hotels from a large number of cities. Furthermore,
we ensured that all reviews in our set had ratings for each of 6 aspects: check-in,
service, value, location, rooms, and cleanliness. The reviews were automatically sentence
split and tokenized.

The multi-aspect rater we used was the PRanking algorithm \cite{crammer2002}, which
is a perceptron-based online learning method. The PRanking algorithm scores each
input feature vector $x \in \mathbb{R}^m$ with a linear classifier,
\[
score_i(x) = w_i \cdot x
\]
Where $score_i$ is the score and $w_i$ the parameter vector for the $i^{th}$ aspect.
For each aspect, the PRanking model also maintains $k$-$1$ boundary
values $b_{i,1},\ldots,b_{i,k-1}$ that
divides the scores into $k$ buckets, each representing a particular rating.
For aspect $i$ a text gets the $j^{th}$ rating if and only if
\[
b_{i,j-1} < score_i(x) < b_{i,j}
\]
Parameters and boundary values are updated using a perceptron style
online algorithm. We used the Snyder and Barzilay
implementation\footnote{http://people.csail.mit.edu/bsnyder/naacl07/}
that was used in their study on agreement models for aspect ranking \cite{snyder2007}.

The input vector $x$ is typically a set of binary features representing
textual cues in a review. Our base set of features are unigram,
bigram and frequently occurring trigrams in the text.
To add topic model features to the input representation we first
estimated the topic distributions for each sentence using both LDA and
MG-LDA. For MG-LDA we could use 
estimators (\ref{ExprTopicDistrEstimGl}) and (\ref{ExprTopicDistrEstimLoc}), but
there is no equivalent estimators for LDA. Instead for both models we set the
probability of a topic for a sentence to be proportional to the number of words 
assigned to this topic. To improve the reliability of the estimator  we
produced 100 samples for each document while keeping assignments of the topics to
all other words in the collection fixed. The probability estimates were then obtained by
averaging over these samples. This approach allows for more direct comparison of
both models. Also, unlike estimators given in (\ref{ExprTopicDistrEstimGl}) and (\ref{ExprTopicDistrEstimLoc}), 
it is applicable to arbitrary text fragments, not necessarily sentences,  which is desirable for topic segmentation. 
We  then found top 3 topic for each sentence using both models, bucketed these topics by their 
probability and concatenated them
with original features in $x$. For example, if a sentence is about
topic 3 with probability between 0.4 and 0.5 and the sentence contains the word ``great'',
then we might have the binary feature
\begin{center}
	$x$ contains ``great'' \& topic=3 \& bucket=0.4-0.5
\end{center}
To bucket the probabilities produced by LDA and MG-LDA we choose
5 buckets using thresholds to distribute the values as evenly as possible.
We also tried many alternative methods for using the real value topic
probabilities and found that bucketing with raw probabilities worked best. Alternatives attempted
include: using the probabilities directly as feature values; normalizing values to (0,1)
with and without bucketing; using log-probabilities with and without bucketing; using
z-score with and without bucketing.

\subsubsection{Results}

All system runs are evaluated using
ranking loss \cite{crammer2002,snyder2007} which measures the average distance between
the true and predicted numerical ratings. If given $N$ test instances, the ranking loss
for an aspect is equal to
\[
\sum_n \frac{|\mbox{actual\_rating}_n - \mbox{predicted\_rating}_n|}{N}
\]
Overall ranking loss is simply the average over each aspect. Note
that a lower loss means a better performance.

We compared four models. The baseline simply rates each aspect as a 5, which is
the most common rating in the data set for all aspects. The second model
is the standard PRanking algorithm over input features, which we denote by ``PRank''. The third model
is the PRanking algorithm but including features derived from the LDA topic model, which is
denoted by ``PRank+LDA''.
The fourth and final model uses the PRanking algorithm but with features derived
from the MG-LDA topic model, which is denoted by ``PRank+MG-LDA''. All topic models
were run to generate 15 topics.

We ran two experiments. The first experiment used only unigram features plus LDA and MG-LDA
features. Results can be seen in \tabref{tab:quant_results}. Clear gains are to be
had by adding topic model features. In particular, the MG-LDA features result in a
statistically significant improvement in loss over using the LDA features. Significance
was tested using a paired t-test over multiple runs of the classifier on different
splits of the data. Results that are significant with a value of $p < 0.001$ are
given in bold. Our second experiment used the full input feature space (unigrams, bigrams,
and frequent trigrams) plus the LDA and MG-LDA features. In this experiment we
would expect the gains from topic model features to be smaller due to the
bigram and trigram features capturing some non-local context, which in fact does happen.
However, there are still significant improvements in performance by adding the MG-LDA
features. Furthermore, the PRank+MG-LDA model still out performs the PRank+LDA model
providing more evidence that the topics learned by multi-grain topic models are
more representative of the ratable aspects of an object.

\begin{table*}
\centering
\caption{Multi-aspect ranking experiments with the PRanking algorithm
for hotel reviews.}
\ \\
\textit{Unigram features only}\\
\begin{tabular}{|l|c|c|c|c|c|c|c|} \hline
Model & Overall & Check-in & Service & Value & Location & Rooms & Cleanliness \\ \hline
Baseline & 1.118 & 1.126 & 1.208 & 1.272 & 0.742 & 1.356 & 1.002 \\
PRank & 0.774 & 0.831 & 0.799 & 0.793 & 0.707 & 0.798 & 0.715 \\
PRank + LDA & 0.735 & 0.786 & 0.762 & 0.749 & 0.677 & 0.746 & 0.690 \\
PRank + MG-LDA & \textbf{0.706} & \textbf{0.748} & \textbf{0.731} & \textbf{0.725} & \textbf{0.635} & \textbf{0.719} & \textbf{0.676} \\
\hline\end{tabular}
\ \\\ \\\ \\
\textit{Unigram, bigram and trigram features}\\
\begin{tabular}{|l|c|c|c|c|c|c|c|} \hline
Model & Overall & Check-in & Service & Value & Location & Rooms & Cleanliness \\ \hline
PRank & 0.689 & 0.735 & 0.725 & 0.710 & 0.627 & 0.700 & 0.637 \\
PRank + LDA & 0.682 & 0.728 & 0.717 & 0.705 & 0.620 & 0.684 & 0.637 \\
PRank + MG-LDA & \textbf{0.669} & \textbf{0.717} & \textbf{0.700} & \textbf{0.696} & \textbf{0.607} & \textbf{0.672} & 0.636 \\
\hline\end{tabular}
\label{tab:quant_results}
\end{table*}

When analyzing the results we can note that for the TripAdvisor data
the MG-LDA model produced clear topics for the \textit{check-in},
\textit{location}, and several coherent \textit{rooms} aspects. This corresponds rather closely with
the improvements that are seen over just the PRank system alone.
Note that we still see an improvement in \textit{service}, \textit{cleanliness} and \textit{value}
since a users ranking of different aspects is highly correlated \cite{snyder2007}.
In particular, users who have favorable opinions of most of the aspects almost
certainly rate \textit{value} high. The LDA model produced clear topics that
correspond to \textit{check-in}, but noisy topics for \textit{location} and \textit{rooms}
with location topics often specific to a single locale (e.g., Paris) and room topics
often mixed with service, dining and hotel lobby terms.

\section{Related Work}
\label{SecRelatedWork}

Recently there has been a tremendous amount of work on summarizing sentiment \cite{beineke2003}
and in particular summarizing sentiment by extracting and aggregating sentiment over ratable aspects. There have been many methods proposed from unsupervised to fully supervised systems.

In terms of unsupervised aspect extraction, in which this work can be categorized,
the system of Hu and Liu \cite{liu2004a,liu2004b}
was one of the earliest endeavors. In that study association mining is
used to extract product aspects that can be rated. Hu and Liu defined an aspect as
simply a string and there was no attempt to cluster or infer aspects that are
mentioned implicitly, e.g., ``The amount of stains in the room was overwhelming''
is about the \textit{cleanliness} aspect for hotels. A similar work by Popescu and Etzioni \cite{popescu2005}
also extract explicit aspects mentions without describing how implicit mentions are extracted and clustered.\footnote{Though they imply that this is done somewhere in their system.}
Clustering can be of particular importance for domains in which aspects
are described with a large vocabulary, such as \textit{food} for restaurants
or \textit{rooms} for hotels. Both implicit mentions and
clustering arise naturally out of the topic model formulation requiring no additional
augmentations.

Gamon et al.\ \cite{gamon2005} present an unsupervised system that does
incorporate clustering, however, their method clusters sentences and not individual aspects
to produce a sentence based summary. Sentence clusters are labeled
with the most frequent non-stop word stem in the cluster.
Carenini et al.\ \cite{carenini2005} present a weakly supervised model
that uses the algorithms of Hu and Liu \cite{liu2004a,liu2004b} to extract
explicit aspect mentions from reviews. The method is extended through
a user supplied aspect hierarchy of a product class. Extracted aspects
are clustered by placing the aspects into the hierarchy using various string
and semantic similarity metrics. This method is then used to compare
extractive versus abstractive summarizations for sentiment \cite{carenini2006}.

There has also been some studies of supervised aspect extraction methods.
For example, Zhuang et al.\ \cite{zhuang2006} work on sentiment summarization 
for movie reviews. In that work, aspects are extracted and clustered, but they are done so manually through the
examination of a labeled data set. The short-coming of such an approach is that
it requires a labeled corpus for every domain of interest.

A key point of note is that our topic model approach is orthogonal
to most of the methods mentioned above. For example, the topic model can be used to help
cluster explicit aspects extracted by \cite{liu2004a,liu2004b,popescu2005} or
used to improve the recall of knowledge driven approaches that require domain
specific ontologies \cite{carenini2005} or labeled data \cite{zhuang2006}.

A closely related model to ours is that of Mei et al. \cite{mei2007} which
performs joint topic and sentiment modeling of collections.  Their
Topic-Sentiment Model (TSM) is essentially equivalent to the PLSA aspect model with 
two additional topics.\footnote{Another difference from PLSA is that Mei et al.\ use a 
background component  to capture common English words.}
One of these topics has a prior towards positive sentiment words, another -
towards negative sentiment words, where both priors are induced from sentiment
labeled data. Though results on  web-blog posts are encouraging,
it is not clear if their method can model sentiments towards discovered topics:
induced distributions of the sentiment words are universal and independent of topics, and
their model uses the bag-of-words assumption, which does not permit exploitation of
co-occurrences of sentiment words with topical words. 
Also it is still not known whether their model can achieve good results on review data, 
because, as discussed in
section~\ref{SectTopicModels} and confirmed in the empirical experiments,
modeling co-occurrences at the document level is
not sufficient. Very recently another approach for joint sentiment and topic
modeling was proposed in~\cite{blei08}. They propose a supervised LDA (sLDA) model which tries
to infer topics appropriate for use in a given classification or regression problem. 
As an application they consider prediction of the overall document  sentiment, 
though they do not consider multi-aspect ranking. 
Both of these joint sentiment-topic models are 
orthogonal to the multi-grain model proposed in our paper. It should be easy to
construct a sLDA or TSM model on top of the MG-LDA model instead of
the standard aspect model. In our work we assumed a sentiment classifier as
a next model in a pipeline, but building a joint sentiment-topic model is
certainly a challenging next step for work.

Several models have
been proposed to overcome the bag-of-words assumption 
by explicitly modeling topic transitions \cite{blei01,griffiths04b,wang05,wallach06,purver06,gruber2007}. 
In our MG-LDA model we instead proposed a sliding windows to model local topics, as it is
computationally less expensive and leads to good results.\footnote{The model
of Blei and Moreno~\cite{blei01} also uses windows, but their windows are not
overlapping and, therefore, it is a priori known from which window a word is going to be
sampled. They perform explicit modeling of topic transitions between these
windows. In our case the distribution of sentences over overlapping windows
$\psi$ is responsible for modeling transitions.}
However, it is possible to construct a multi-grain model which 
uses a n-gram topic model for local topics and a distribution fixed per
document for global topics.

\section{Summary and Future Work}

In this work we presented multi-grain topic models and showed that they are
superior to standard topic models when extracting ratable aspects from
online reviews. These models are particularly suited to this problem since
they not only identify important terms, but also cluster them into
coherent groups, which is a deficiency of many previously proposed
methods.

There are many directions we plan on investigating in the future
for the problem of aspect extraction from reviews.
A promising possibility is to develop a supervised version of the
model similar to supervised LDA \cite{blei08}. In such a model
it would be possible to infer topics for a multi-aspect classification
task. There are many data sets available that could be used, most
notably the TripAdvisor data that was used as part of this study.
Another direction would be to investigate hierarchical topic models.
Ideally for a corpus of restaurant reviews, we could induce a hierarchy
representing cuisines. Within each cuisine we could then extract cuisine specific
aspects such as \textit{food} and possibly \textit{decor} and \textit{atmosphere}.
Other ratable aspects like \textit{service} would ideally be shared across all cuisines
in the hierarchy since there typically is a standard vocabulary for describing them.

The next major step in this work is to combine the aspect extraction methods
presented here with standard sentiment analysis algorithms to aggregate and
summarize sentiment for products and services. Currently we are
investigating a two-stage approach where aspects are first extracted and sentiment
is then aggregated. However, we are also interested in examining joint models
such as the TSM model~\cite{mei2007}.


\bibliographystyle{abbrv}
\bibliography{sigproc}  

\end{document}